\documentclass[11pt]{article}

\usepackage[final]{acl}

\usepackage{times}
\usepackage{latexsym}

\usepackage[T1]{fontenc}
\usepackage[utf8]{inputenc}

\usepackage{microtype}

\usepackage{inconsolata}

\usepackage{graphicx}

\usepackage{amsmath}
\usepackage{amsfonts}
\usepackage{amssymb}
\usepackage{multirow}
\usepackage{tabularx}
\usepackage{booktabs}
\usepackage{pifont}
\title{ImmersiveTTS: Environment-Aware Text-to-Speech with Multimodal Diffusion Transformer and Domain-Specific Representation Alignment}

\author{
    Jun-Hak Yun, Seung-Bin Kim, 
    Seong-Whan Lee\thanks{Corresponding author}  \\
    Department of Artificial Intelligence, Korea University, Seoul, Korea\\
    \texttt{\{jh\_yun, sb-kim, sw.lee\}@korea.ac.kr}
    }

\begin{document}
\maketitle
\begin{abstract}
Recent advancements in text-guided audio generation have yielded promising results in diverse domains, including sound effects, speech, and music. 
However, jointly generating speech with environmental audio remains challenging due to the inherent disparities in their acoustic patterns and temporal dynamics.
We propose ImmersiveTTS, an environment-aware text-to-speech (TTS) model that generates natural speech seamlessly integrated within environmental contexts by explicitly modeling cross-modal interactions.
Our model builds on a multimodal diffusion transformer and fuses transcript-aligned speech latent with text-conditioned environmental context via joint attention.
To enhance semantic consistency, we introduce a domain-specific representation alignment objective tailored to environment-aware TTS, leveraging complementary self-supervised representations from speech and audio encoders.
Experimental results show that ImmersiveTTS achieves higher naturalness, intelligibility, and audio fidelity than existing approaches across objective metrics and human listening tests.
\end{abstract}

\section{Introduction}
Text-guided audio generation has emerged as a prominent research area in speech and audio processing, broadly categorized into sub-tasks: text-to-audio (TTA) and text-to-speech (TTS).
TTA \citep{liu2023audioldm,liu2024audioldm,kreuk2022audiogen,huang2023make} focuses on synthesizing non-speech audio, including foley effects, music, and environmental soundscapes, from natural language descriptions.
Because most TTA models are not designed or optimized to capture fine-grained phonetic and prosodic structure, they often struggle to synthesize intelligible speech with precise linguistic content, even when given instructions such as “A woman is speaking.”

In contrast, TTS \citep{ren2020fastspeech,kim2021conditional,lee2022hierspeech, lee2025hierspeech++, kim2025fillerspeech, chen2025neural} aims to generate natural-sounding human speech waveforms from textual input, such as characters, phonemes, or words.
Despite its success, robust speech generation across diverse acoustic environments remains a significant challenge in TTS research.
This difficulty stems from two main factors: (i) speech and environmental audio have largely been modeled in separate pipelines; and (ii) jointly generating heterogeneous audio sub-modalities, such as intelligible speech with environmental audio, within a single model is inherently complex, due to substantial differences in their acoustic structures and modeling requirements.

Motivated by these challenges, several recent studies have explored unified modeling for multiple audio generation tasks \citep{vyas2023audiobox,yang2023uniaudio,liu2024audioldm,choi2024dddm}. 
These approaches leverage powerful generative backbones such as diffusion models \citep{ho2020denoising,song2020score,rombach2022high}, conditional flow matching \citep{lipman2022flow,liu2022flow, yun2025flowhigh}, and language models \citep{chen2025neural}.
Nevertheless, many of these systems optimize for each task separately and still struggle to synthesize natural speech with well-integrated environmental audio.

In particular, environment-aware TTS \citep{lee2024voiceldm,jung2025voicedit} has been explored as a method for generating speech and its surrounding acoustic context simultaneously.
However, existing methods do not fully capture cross-modal interactions between speech and environmental audio.
As a result, the synthesized outputs often exhibit speech-environment mismatch, leaving substantial room for improvement in overall coherence and immersion.

In this paper, we propose ImmersiveTTS, an environment-aware TTS model that jointly synthesizes natural speech with environmental audio by explicitly modeling interactions between linguistic content and environmental context, thus addressing these limitations. 
To enable this, we build on the multimodal diffusion transformer (MM-DiT) architecture \citep{esser2403scaling}, which was originally designed for image-text integration with a dual-stream backbone.
We extend this approach to speech synthesis.
Specifically, we assign transcript-aligned speech features and text-conditioned environmental context to parallel streams.
We use joint attention between the two streams to explicitly model cross-modal interactions.

Although the MM-DiT architecture supports multimodal joint training, relying solely on its generative objective may be insufficient to learn speech representations that are both linguistically precise and grounded in environmental context.
To stabilize cross-modal learning and improve semantic consistency, we introduce a domain-specific representation alignment objective tailored to environment-aware TTS, inspired by representation alignment (REPA) \citep{yu2024representation}.
Experimental results show that ImmersiveTTS achieves higher naturalness, intelligibility, and speech-environment coherence than existing methods.
Ablation studies further validate the effectiveness of domain-specific alignment for environment-aware TTS.
Audio samples and code implementations are provided at \url{https://jjunak-yun.github.io/ImmersiveTTS}.

\section{Related Work}

\subsection{Environment-Aware Text-to-Speech}
Environment-aware TTS aims to generate speech that matches a target acoustic environment, such as a background noise condition or a sound scene.
Existing approaches can be organized according to how they obtain environmental information.

The first type of methods \citep{tan22_interspeech,lu2025incremental,glazer2025umbratts,lu2025daien} infer the acoustic context from reference audio and condition the TTS system on speaker and environmental attributes derived from it, either encoded as embeddings or used directly.
In particular, \citep{tan22_interspeech} extends a Tacotron \citep{shen2018natural} by introducing separate encoders for speaker identity and room acoustics, while IDEA-TTS \citep{lu2025incremental}, based on VITS \citep{kim2021conditional}, incrementally disentangles speaker, content, and environment factors from reference speech.
Building on a flow matching TTS backbone \citep{chen-etal-2025-f5}, UmbraTTS \citep{glazer2025umbratts} introduces speech-to-environment ratio conditioning, 
while DAIEN-TTS \citep{lu2025daien} uses a pretrained speech-environment separation module for environment-aware TTS.

The second type of methods \citep{lee2024voiceldm,jung2025voicedit} take natural language prompts as input to describe the target acoustic scene, rather than relying on reference audio.
Extending the AudioLDM framework, VoiceLDM \citep{lee2024voiceldm} conditions a U-Net on two natural language prompts.
A description prompt is encoded by a frozen CLAP encoder \citep{wu2023large}, while a content prompt is encoded by a SpeechT5 encoder \citep{ao2022speecht5}. 
The resulting embeddings are injected into the U-Net via cross-attention.
More recently, VoiceDiT \citep{jung2025voicedit} adopts Diffusion transformers (DiTs) with adaptive layer normalization (AdaLN) for environmental conditioning, enabling environment-aware speech synthesis from both text and visual prompts.
Related studies, such as ViT-TTS \citep{liu2023vit} and M2SE-VTTS \citep{liu2025multi}, also explore visual or spatial cues as additional modalities for TTS.

Compared with reference audio-based approaches, specifying the environment via text prompts offers advantages: it scales more naturally to arbitrary or unseen acoustic scenes and obviates the need to collect reference recordings.
In this work, we focus on the latter strategy: utilizing text prompts to directly specify the desired environmental context.
Despite these advantages, effectively fusing distinct textual cues, namely speech transcriptions and environmental descriptions, into a seamlessly integrated audio waveform remains a non-trivial modeling challenge.

\subsection{Multimodal Diffusion Transformers}
DiTs \citep{peebles2023scalable} have been introduced as scalable alternatives to conventional U-Net backbones \citep{ronneberger2015u} in diffusion models.
The MM-DiT architecture proposed in SD3 \citep{esser2403scaling} extends DiT to the multimodal setting by mapping text tokens and image patches into a unified token sequence and applying self-attention over all tokens.
This approach allows for bidirectional cross-modal interaction at every layer within a single transformer.
To accommodate modality-specific properties, MM-DiT adopts a dual-stream design that maintains separate representation paths for image and text tokens.
In addition, multiple text encoders, such as CLIP \citep{radford2021learning} and T5 \citep{raffel2020exploring}, are supported within this unified design.

Following these advancements, recent audio generative models \citep{fei2024flux,hung2024tangoflux,li2025meanaudio,cheng2025mmaudio,liu2025thinksound,wang2025kling,shan2025hunyuanvideo} have adopted MM-DiT to condition audio generation on textual prompts and other contextual information in the latent space.
These developments underscore the flexibility and effectiveness of MM-DiT as a backbone for multimodal audio generation.
In this work, we adapt the MM-DiT architecture for environment-aware TTS. Unlike previous general audio models, we specialize its dual-stream design to treat transcript-aligned speech features and environmental cues as distinct yet interacting modalities, thereby facilitating precise linguistic control within immersive acoustic scenes.

\begin{figure*}[!t]
    \centerline{\includegraphics[width=0.99\textwidth]{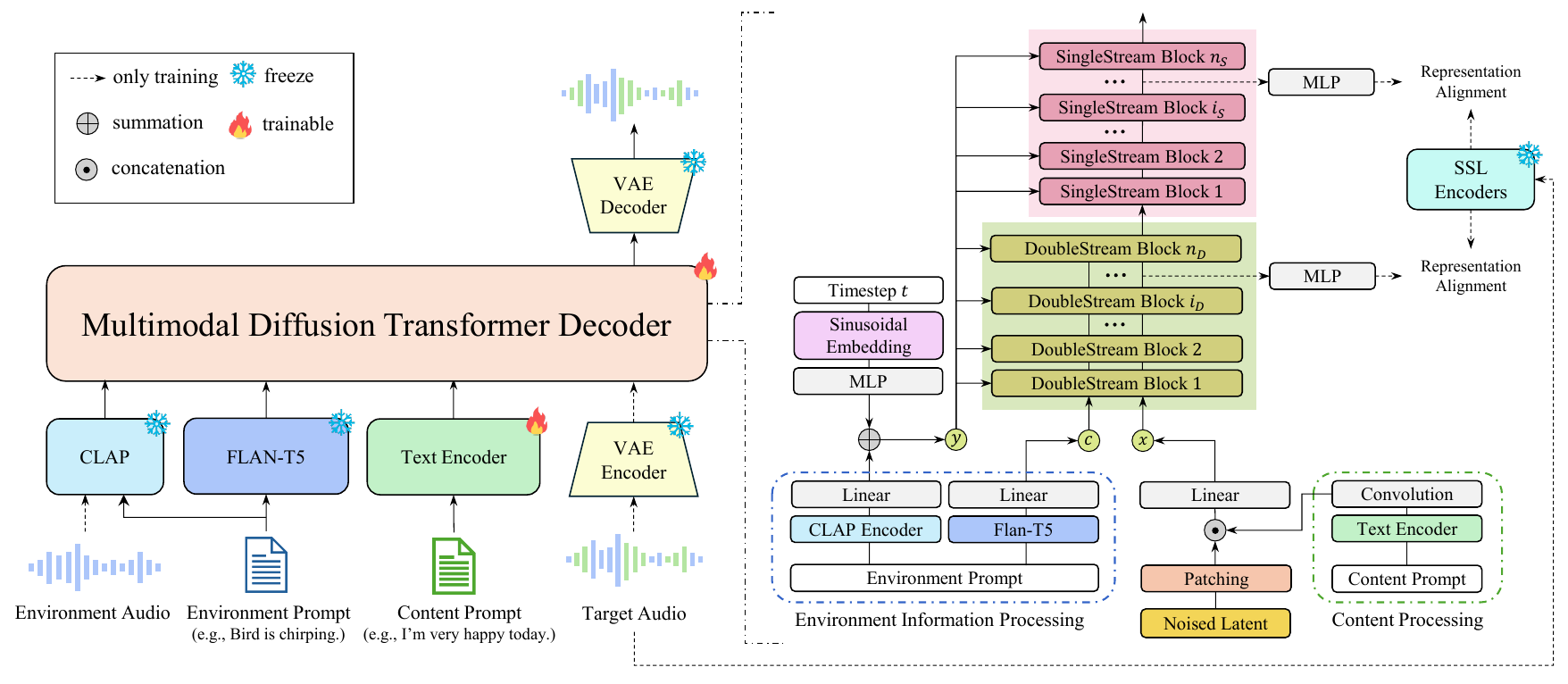}} \vspace{-0.3cm}
    \caption{Overview of ImmersiveTTS. A dual-stream MM-DiT backbone conditions the speech stream on content prompt-aligned linguistic features. At the same time, Flan-T5 token embeddings drive the environmental context stream, and CLAP embeddings modulate AdaLN for global conditioning. The model is trained with flow matching and domain-specific REPA objectives.}
    \label{fig1:overall} 
    \vspace{-0.2cm}
\end{figure*}

\subsection{Representation Alignment}
The REPA method \citep{yu2024representation} regularizes diffusion transformers by aligning intermediate hidden states of DiT with the features produced by a powerful self-supervised learning (SSL) teacher encoder \citep{oquab2023dinov2}.
It is designed to improve semantic fidelity and accelerate convergence in diffusion and flow matching models.

Although REPA was first introduced in the context of image generation, subsequent work has begun to adopt REPA-based objectives for TTS and TTA tasks.
ACE-Step \citep{gong2025ace} incorporates a semantic alignment loss, aligning intermediate features from its Linear DiT \citep{xie2024sana} with representations from pretrained MERT \citep{li2023mert} and mHuBERT \citep{boito2024mhubert}.
Vevo2 \citep {zhang2025vevo2} adopts REPA in its flow matching acoustic model, aligning an intermediate representation with W2v-BERT 2.0 features \citep {chung2021w2v} to improve training efficiency and controllability of speech and singing voice generation.
A-DMA \citep{choi2025accelerating} introduces text and speech-guided alignment losses using a CTC \citep{graves2006connectionist} and a speech SSL model such as HuBERT \citep{hsu2021hubert}, and shows that these alignment objectives accelerate convergence and improve speech quality.
Building on these insights, our approach employs a domain-specific alignment scheme that uses separate pretrained SSL encoders to capture the distinct properties of speech and the environment. 
We elaborate on this architectural design in the following section.

\section{ImmersiveTTS}

In this section, we present ImmersiveTTS, an environment-aware TTS model built on the MM-DiT backbone to capture the interplay between speech and environmental context. 
For high-fidelity generation and stable training, we adopt a flow matching generative objective coupled with a domain-specific REPA.
The overall pipeline is illustrated in Figure~\ref{fig1:overall}, and the details of each component are described below.

\subsection{Preliminaries on Flow Matching}
Flow matching or rectified flow \citep{lipman2022flow, liu2022flow} provides an approach that aims to learn a transformation between simple prior $\pi_0$ and data distribution $\pi_1$ on $\mathbb{R}^{d_z}$.
The transformation is expressed as the following ordinary differential equation (ODE) over time $t\in [0,1]$:
\begin{equation}
    \label{eq:flow}
    \frac{\text{d}}{\text{d}t} Z_t= v(Z_t,t), \quad Z_0 \sim \pi_0, \quad Z_1 \sim \pi_1,
\end{equation}
where $v : \mathbb{R}^{d_z} \times [0,1] \to \mathbb{R}^{d_z}$ is the time-dependent velocity field and $\pi_0$ typically follows a standard Gaussian distribution $\mathcal{N} (0, I)$. 

We parameterize the field with a neural network $v_\theta$.
It is trained by minimizing the mean squared error between the velocity of straight paths connecting random pairs $(Z_0, Z_1)$ and the neural velocity as follows:
\begin{equation}
    \label{eq2:flowloss}
    \mathcal{L}_{\text{Flow}}(\theta) = \mathbb{E}_{t,Z_0, Z_1} \Big[|| (Z_1 - Z_0) - v_\theta(Z_t, t)||^2\Big],
\end{equation}
where $Z_t = (1-t)Z_0 + tZ_1 $ represents the linear interpolation between $Z_0 \sim \pi_0$ and $Z_1 \sim \pi_1$, and $t\in [0,1]$ denotes the time step.
Given the learned velocity field $v_\theta$, the flow-based model transports samples from the prior $\pi_0$ to the target distribution $\pi_1$ along straight trajectories.

\subsection{Audio Compression}
\label{ssec:audiocompression}
To capture both speech and general audio characteristics within a unified latent space, we employ the pretrained variational autoencoder (VAE) used in AudioLDM2 \citep{liu2024audioldm}.
Let $X_{\text{wav}} \in \mathbb{R} ^ {d \cdot f_s}$ denote the raw waveform of duration $d$ seconds with a sampling rate of $f_s$.
We first convert $X_{\text{wav}}$ into a log-mel spectrogram $X_{\text{mel}} \in \mathbb{R}^{F \times L}$, where $F$ is the number of mel bins and $L$ is the length of mel-spectrogram sequence.
The VAE encoder compresses $X_{\text{mel}}$ into a latent representation $Z \in \mathbb{R}^{8 \times F/4 \times L/4}$ by downsampling the time-frequency axes by a factor of 4.
The VAE decoder reconstructs $\hat{X}_{\text{mel}}$ from $Z$, and a pretrained vocoder \citep{kong2020hifi, kim2021fre} converts $\hat{X}_{\text{mel}}$ into the waveform $\hat{X}_{\text{wav}}$.
All VAE parameters are frozen during training.

\subsection{Multimodal Diffusion Transformer for Environment-Aware Text-to-Speech}
Our objective is to generate speech that simultaneously preserves linguistic content and aligns with an environmental context. 
Accordingly, the model is conditioned on two distinct textual inputs: a content prompt $y_\text{cont}$ (i.e., the transcription) and an environment prompt $y_\text{env}$ (i.e., the background description). 
To effectively model the interplay between speech latents and environmental cues, we employ the MM-DiT backbone \citep{esser2403scaling}, which accommodates heterogeneous inputs and provides a robust foundation for cross-modal fusion.

In particular, we adopt the Flux architecture\footnote{\url{https://github.com/black-forest-labs/flux}} to synthesize high-fidelity speech.
The architecture comprises a stack of double-stream DiT layers followed by single-stream DiT layers.
In the double-stream stage, we decouple the processing into two dedicated pathways: 
(i) the environmental context stream, which encodes the fine-grained environmental context tokens derived from $y_\text{env}$, and (ii) the speech stream, which processes the noisy audio latents $Z_t$ conditioned on the linguistic features from $y_\text{cont}$. 
Crucially, these parallel streams exchange information through joint attention mechanisms. 
The detailed internal flow of the double-stream DiT block is illustrated in Appendix~\ref{appendix:implementation}.
This design allows the speech generation process to dynamically attend to and harmonize with the environmental cues without losing its linguistic structure. 
Subsequently, only the representations from the speech stream are forwarded to the single-stream blocks, where they are further refined via self-attention layers.
We describe the detailed configuration of each stream below.\\[8pt]
% \subsubsection{Environment Context Stream}
\textbf{Environmental Context Stream}. 
For audio generation, existing approaches typically rely on either CLAP \citep{wu2023large} encoders for coarse, global sound semantics or T5-family \citep{raffel2020exploring} encoders for fine-grained detail. 
We adopt a dual-granularity conditioning strategy that leverages the complementary strengths of both the CLAP and T5 encoders \citep{xue2024auffusion}.

First, to capture the global acoustic context, we project the CLAP embedding from $y_\text{env}$ using an MLP and combine it with the diffusion timestep embedding to condition the AdaLN modules.
By modulating the AdaLN scale and shift parameters ($\gamma, \beta$) across transformer blocks, it globally conditions the generation process.

In parallel, we apply a linear projection to token-level T5 embeddings of $y_\text{env}$ to match the model dimension, and feed them into the environment context stream as an input sequence.
This allows the speech stream to selectively attend to local environmental details through the joint attention mechanism in the double-stream layers.
This approach balances global semantic consistency with fine-grained acoustic fidelity.\\[8pt]
% \subsubsection{Environment-Aware Speech Stream}
\textbf{Environment-Aware Speech Stream}.
To synthesize intelligible speech that faithfully follows the content prompt $y_\text{cont}$, we incorporate an explicit temporal alignment that directly injects linguistic features into the speech stream.
Following the framework of \citep{kim2020glow}, the text encoder converts $y_\text{cont}$ into a hidden representation $\tilde{\mu}_{1:L}$, while the monotonic alignment search (MAS) algorithm estimates phone-level durations $d'_{1:L}$. The hidden vectors are then expanded based on $d'$ to produce a frame-level prior mel representation $\mu$.
The text encoder and duration predictor are mainly optimized using the prior loss $\mathcal{L}_{\text{Prior}}$ and MAS-based duration loss $\mathcal{L}_{\text{Dur}}$ as in \citep{kim2020glow}.

To align the prior representation $\mu$ with the audio latent space, $\mu$ is processed through a convolution network, which bridges the structural gap between the mel-spectrogram space and the VAE latent manifold \citep{jung2025voicedit}.
The resulting features are concatenated with the noisy latent $Z_t$ along the channel dimension and fed into the environment-aware speech stream.
Within the MM-DiT layers, this speech stream actively exchanges information with the sequence of the environment context stream via joint attention.
After passing through the full stack of double-stream blocks, only the environmentally-adapted speech representations are forwarded to the single-stream blocks for high-fidelity refinement.

\subsection{Domain-Specific Representation Alignment}
Without explicit feature-level alignment during training, we find that the diffusion backbone often struggles to simultaneously preserve linguistic intelligibility and environmental fidelity along the denoising trajectory, leading to content errors and acoustically inconsistent scenes.
To enhance training stability and convergence, we extend the REPA strategy \citep{yu2024representation} to our multi-domain setting involving speech and environmental audio.\\[8pt]
%\subsubsection{Domain-Specific SSL Encoders}
\textbf{Domain-Specific SSL Encoders}.
For domain-specific REPA, we adopt a dual-teacher strategy that leverages the complementary strengths of specialized encoders.
Building on the insights of \citep{chang2025usad}, we use WavLM \citep{chen2022wavlm} and ATST-Frame \citep{li2024self} as target encoders: (i) WavLM, a speech-specialized SSL model, selected to enforce precise phonetic and linguistic fidelity; and (ii) ATST-Frame, an audio-specialized SSL model, chosen to capture rich environmental acoustic events.
Aligning to this heterogeneous pair rather than a single encoder encourages target representations that reflect both high-fidelity linguistic content and detailed environmental context.\\[8pt]
%\subsubsection{Alignment Objective}
\textbf{Alignment Objective}.
Let $\{E_k\}_{k=1}^K$ denote a set of $K$ pretrained SSL encoders.
For a target audio input $X\sim p_\text{data}$, the $k$th encoder yields a target representation $r_k = E_k (X)\in \mathbb{R}^{B\times L_k \times D_k}$, where $B$ denotes the batch size, and $L_k$, $D_k$ represent the sequence length and dimensionality, respectively.
To align our model with these targets, we extract hidden features $h_k \in \mathbb{R}^{B \times L_h \times D_h}$ specifically from the intermediate layers of the speech stream.
These features are passed through a lightweight MLP projector to obtain $h'_k = \text{MLP}_k(h_k) \in \mathbb{R}^{B \times L_h \times D_k}$, mapping the transformer features into the encoder representation space.
Following \citep{gong2025ace}, we match the temporal resolutions of the projected features $h'_k$ and target features $r_k$ by interpolating or pooling them to a common temporal length $\tilde{L}$, yielding synchronized sequences $\tilde{h}'_k$ and $\tilde{r}_k$.
The REPA loss is based on cosine similarity $\mathrm{CosSim}(\cdot,\cdot)$ defined as follows:
\begin{equation}
    \mathcal{L}_{\text{SSL}_k} = - \mathbb{E}_{X} \left[ \mathrm{CosSim} (\tilde{r}_{k}, \tilde{h}'_k)  \right].
\end{equation}
Finally, the total objective is a weighted sum of the domain-specific alignment losses:
\begin{equation}
    \mathcal{L}_{\text{REPA}} 
    = \sum_{k=1}^K \lambda_{k} \mathcal{L}_{\text{SSL}_k},
\end{equation}
where $\lambda_{k}$ is a hyperparameter to control the influence of each teacher.
In our experiments, we set $\lambda_{k} = 1$ for all $k$.

\subsection{Training and Inference}
\textbf{Training}. 
The model is optimized with four losses during training. 
The velocity predictor and convolutional mapper are optimized with the flow matching objective $\mathcal{L}_\text{Flow}$ and the alignment objective $\mathcal{L}_\text{REPA}$.
The text encoder and duration predictor receive gradients backpropagated through the conditioning pathway and, in addition, are directly supervised by the MAS-based prior loss $\mathcal{L}_\text{Prior}$ and the duration loss $\mathcal{L}_\text{Dur}$.
Our final objective is 
\begin{equation}
    \mathcal{L}
    = \lambda_{\text{P}} \mathcal{L}_\text{Prior} 
    + \lambda_{\text{D}} \mathcal{L}_\text{Dur} 
    + \lambda_{\text{F}} \mathcal{L}_\text{Flow}
    + \lambda_{\text{R}} \mathcal{L}_\text{REPA},
\end{equation}
where we set all loss weights to 1 in our experiments.
We freeze the CLAP and T5 encoders and draw the timestep $t\in(0,1)$ from a logit-normal distribution with mean of 0 and variance of 1 \citep{esser2403scaling}, rather than uniformly from $\mathrm{U}(0,1)$.

To enable flexible control over synthesized attributes, we adopt dual classifier-free guidance (CFG) \citep{ho2022classifier,lee2024voiceldm} by independently masking the content and environment prompt sequences with probability $0.1$ during training.\\[8pt]
\textbf{Inference}.
During sampling, we first sample a random noise $Z_0 \sim \mathcal{N}(\mathbf{0},\mathbf{I})$.
The explicit velocity field is adjusted using dual CFG as follows:
\begin{align}
    \nonumber&\tilde{v}_\theta(Z_t,y_{\mathrm{env}},y_{\mathrm{cont}} ) = v_\theta(Z_t,y_{\mathrm{env}},y_{\mathrm{cont}})\\\nonumber 
    &+ \omega_{\mathrm{env}} \Big( v_\theta(Z_t, y_{\mathrm{env}}, \emptyset_{\mathrm{cont}})- v_\theta(Z_t,\emptyset_{\mathrm{env}}, \emptyset_{\mathrm{cont}}) \Big)\\    
    &+ \omega_{\mathrm{cont}} \Big( v_\theta(Z_t, \emptyset_{\mathrm{env}}, y_{\mathrm{cont}})- v_\theta(Z_t,\emptyset_{\mathrm{env}}, \emptyset_{\mathrm{cont}}) \Big),
\end{align}
where $\omega_{\mathrm{env}}$ and $\omega_\mathrm{cont}$ denote the guidance scale of each modality, while $\emptyset_{\mathrm{env}}$ and $\emptyset_{\mathrm{cont}}$ denote the corresponding null conditions.
We then employ Euler's method to solve the ODE in Equation~\ref{eq:flow}:
\begin{equation}
    Z_{t+\tau} = Z_t + \tau \cdot \tilde{v}_\theta (Z_t,t,y_{\mathrm{env}},y_{\mathrm{cont}} ).
\end{equation}
Leveraging flow matching-based ODE sampling, we can generate high-quality latent features in fewer sampling steps.
Finally, we decode the generated latents with the VAE decoder and synthesize waveforms using the pretrained vocoder.

%%%%%%%%%%%%%%%%%%%%%%%%%%%%%%%%%%%%%%
%             TABLE 1  
%%%%%%%%%%%%%%%%%%%%%%%%%%%%%%%%%%%%%%
\begin{table*}
  \centering
  \resizebox{\textwidth}{!}{
  \begin{tabular}{l c c ccc ccc}
    \toprule
    
    & & & \multicolumn{3}{c}{\textbf{Subjective}} & \multicolumn{3}{c}{\textbf{Objective}} \\
    
    \cmidrule(lr){4-6} \cmidrule(lr){7-9}
    \textbf{Model} & \textbf{\#Param.}& \textbf{NFEs} & \textbf{SN-MOS($\uparrow$)} & \textbf{EC-MOS($\uparrow$)} & \textbf{ON-MOS($\uparrow$)} 
    & \textbf{WER($\downarrow$)} & \textbf{FAD($\downarrow$)} & \textbf{CLAP($\uparrow$)} \\
    
    \hline
    Ground Truth             
    & -  & - & - & - & - & 22.29 & - & 0.503 \\ 
    Reconstructed    
    & -  & - & 4.08 $\pm$ 0.08 & 4.16 $\pm$ 0.08 & 3.49 $\pm$ 0.05 & 22.58 & - & 0.488 \\
    \hline
    VoiceLDM \citep{lee2024voiceldm} 
    & 508M & 200 & 3.41 $\pm$ 0.06 & 3.33 $\pm$ 0.07 & 2.55 $\pm$ 0.05 & 16.45 & 8.75  & 0.229  \\
    VoiceDiT \citep{jung2025voicedit} 
    & 566M & 200 & 3.47 $\pm$ 0.05 & 3.44 $\pm$ 0.07 & 2.63 $\pm$ 0.05 & 11.68 & 9.07  & 0.263  \\
    \hline
    ImmersiveTTS
    & 450M & 25 & \textbf{4.20} $\pm$ 0.07 & \textbf{3.48} $\pm$ 0.07 & \textbf{3.47} $\pm$ 0.05 &\textbf{8.06} & \textbf{5.80}   & \textbf{0.308}  \\ 
    \bottomrule
  \end{tabular}
  }
    \vspace{-0.25cm}\caption{\label{table:envtts}
    Experimental results for environment-aware text-to-speech on the AudioCaps \textit{test} set. \textbf{\#Param.} denotes the number of trainable parameters. The MOS results are reported with a 95\% confidence interval.
  }
\end{table*}

%%%%%%%%%%%%%%%%%%%%%%%%%%%%%%%%%%%
%             TABLE 2
%%%%%%%%%%%%%%%%%%%%%%%%%%%%%%%%%%%
\begin{table*}
  \centering
  \resizebox{\textwidth}{!}{
  \begin{tabular}{l c c ccc ccc}
    \toprule
    
    & & & \multicolumn{3}{c}{\textbf{Subjective}} & \multicolumn{3}{c}{\textbf{Objective}} \\
    
    \cmidrule(lr){4-6} \cmidrule(lr){7-9}
    \textbf{Model} & \textbf{\#Param.}& \textbf{NFEs} & \textbf{SN-MOS($\uparrow$)} & \textbf{EC-MOS($\uparrow$)} & \textbf{ON-MOS($\uparrow$)} 
    & \textbf{WER($\downarrow$)} & \textbf{FAD($\downarrow$)} & \textbf{CLAP($\uparrow$)} \\

    \hline
    Ground Truth (Augmented)             
    & -  & - & - & - & - & 7.86 & - & 0.317 \\ 
    Reconstructed    
    & -  & - & 4.02 $\pm$ 0.08 & 3.95 $\pm$ 0.08 & 3.41 $\pm$ 0.07 & 3.59 & - & 0.291 \\
    \hline
    VoiceLDM \citep{lee2024voiceldm} 
    & 508M & 200 & 3.32 $\pm$ 0.06 & 3.24 $\pm$ 0.07 & 2.91 $\pm$ 0.08 & 11.20 & 6.98  & 0.118  \\
    VoiceDiT \citep{jung2025voicedit} 
    & 566M & 200 & 3.45 $\pm$ 0.06 & \textbf{3.38} $\pm$ 0.06 & 3.12 $\pm$ 0.08 & 7.08 & 5.37  & 0.134  \\ 
    \hline
    ImmersiveTTS
    & 450M & 25 & \textbf{4.18} $\pm$ 0.07 & 3.32 $\pm$ 0.06 & \textbf{3.23} $\pm$ 0.08 &\textbf{4.48} & \textbf{3.92}   & \textbf{0.207}  \\ 
    \bottomrule
  \end{tabular}
  } \vspace{-0.2cm}
    \caption{\label{table:envtts2}
    Experimental results for environment-aware text-to-speech on the augmented test set with Seed-TTS \textit{test-en} and AudioCaps \textit{test} sets. The MOS results are reported with a 95\% confidence interval.
  } \vspace{-0.25cm}
  
\end{table*}

\section{Experiments}

\subsection{Experimental Setup}
\textbf{Datasets}.
To construct a robust training corpus for environment-aware TTS, we use two datasets: LibriTTS \citep{zen2019libritts} for high-quality speech and WavCaps \citep{mei2024wavcaps} for diverse environmental sounds. 
We use the \textit{train-clean-360} subset of LibriTTS to provide clean linguistic content.
WavCaps contains 400k audio clips; we explicitly filter out samples containing spoken content, retaining 340k non-speech clips to avoid overlapping speech. 
Following \citep{jung2025voicedit}, we construct the training corpus by mixing clean speech from LibriTTS with environmental sounds from WavCaps.
For each mixture, the environmental audio sample is mixed at a signal-to-noise ratio (SNR) value uniformly sampled between 2 and 10 dB.
To ensure the model maintains the capability to generate clean speech, we skip this mixing process and use clean speech only with probability $0.15$.\\[8pt]
\textbf{Preprocessing}.
All audio samples are downsampled to 16 kHz and converted into a mel-spectrogram with 64 mel bins using an STFT with an FFT size of 1024, window size of 1024, and hop length of 160.
The frozen AudioLDM2\footnote{\url{https://huggingface.co/cvssp/audioldm2}} VAE then encodes this spectrogram into an 8-channel latent representation, which we use as the training target.\\[8pt]
\textbf{Implementation Details}.
ImmersiveTTS is trained for 400k steps on 2 NVIDIA RTX A6000 GPUs using the AdamW optimizer at a constant learning rate of $1\times10^{-4}$, and a batch size of 8 per GPU.
The velocity field estimator consists of 12 double-stream blocks, 18 single-stream blocks, 6 attention heads, and a hidden state dimension of 1024, totaling approximately 450M trainable parameters. Detailed implementation information can be found in Appendix \ref{appendix:implementation}.

\subsection{Evaluation Metrics}
We evaluate the performance of environment-aware TTS using both subjective and objective metrics.
We conduct a mean opinion score (MOS) test to assess three aspects of the generated audio on a 5-point scale (1 to 5):
speech naturalness (SN-MOS), environmental consistency (EC-MOS), and overall integration naturalness (ON-MOS).
Detailed information on the MOS can be found in Appendix~\ref{appendix:mos}.

For objective evaluation,
we employ the word error rate (WER) to assess speech intelligibility and content accuracy.  
For WER, we use Whisper-\textit{Large-v3} \citep{radford2023robust} as the automatic speech recognition (ASR) model.
We additionally measure speaker similarity using speaker embedding cosine similarity (SECS) computed with WavLM-\textit{base-sv} \citep{chen2022wavlm}, which serves as an objective metric for speaker identity preservation.
We report the Frechet audio distance (FAD) to measure the distribution distance between generated and target audio using VGGish \citep{hershey2017cnn}, and the CLAP score to quantify text-audio coherence with the environment description, defined as the cosine similarity between the CLAP embeddings of the text prompt and the synthesized audio.
We also report the number of function evaluations (NFEs) as a measure of sampling efficiency.

\section{Results}

\subsection{Main Results}
\label{ssec:main}
We compare ImmersiveTTS with diffusion-based environment-aware TTS models, VoiceLDM~\citep{lee2024voiceldm} and VoiceDiT~\citep{jung2025voicedit}, which condition generation on natural language environment descriptions via a pretrained CLAP text encoder. 
For a fair comparison, all models are trained from scratch on the same training corpus with the same number of optimization steps.
We also select optimal dual CFG scales based on preliminary tuning on an evaluation set.\footnote{We use $(\omega_{\text{env}}, \omega_{\text{cont}})=(3,5)$ for VoiceLDM and $(5,3)$ for VoiceDiT.}

Table~\ref{table:envtts} reports the performance on the AudioCaps \textit{test} set, where the ground truth audio contains both speech and background audio.
\textit{Reconstructed} denotes the reconstruction of the target audio obtained by encoding and decoding it through the STFT, VAE, and vocoder following Section~\ref{ssec:audiocompression}.
We use \textit{Reconstructed} as the practical upper bound for subjective evaluation.
We note that relatively high WER for ground-truth and reconstructed samples has also been reported in prior research~\citep{lee2024voiceldm,jung2025voicedit,lu2025daien}, as ASR can degrade when background audio partially masks speech.

Compared to existing approaches, ImmersiveTTS achieves substantially higher subjective scores, especially in SN-MOS and ON-MOS, indicating that our model generates more natural speech that is better integrated into the acoustic environment.
ImmersiveTTS also improves objective metrics, yielding lower WER and FAD and higher CLAP score, suggesting better intelligibility, audio quality, and text-audio semantic alignment.
It validates that our joint attention and domain-specific REPA strategy successfully preserves linguistic clarity while ensuring semantic alignment with the environment.

Table~\ref{table:envtts2} further evaluates models on an augmented test set constructed from Seed-TTS \textit{test-en} and non-speech clips from AudioCaps \textit{test} set, where clean speech is mixed with environmental audio.
Although VoiceDiT obtains a slightly higher EC-MOS, ImmersiveTTS achieves the best SN-MOS and ON-MOS, indicating that it simultaneously achieves stronger overall naturalness and speech-environment integration.
ImmersiveTTS also achieves the lowest WER and improves FAD and CLAP compared to existing methods.
Notably, ImmersiveTTS attains these gains with only 25 NFEs, compared to 200 for the diffusion baselines.
Overall, ImmersiveTTS shows consistent improvements in both real and augmented test sets.
Additional evaluation of speaker similarity and broader baseline comparisons are provided in Appendix~\ref{appendix:smos} and Appendix~\ref{appendix:broad_baseline}, respectively.

\begin{table}[t]
  \centering
  \resizebox{\columnwidth}{!}{%
  \begin{tabular}{l | c | ccc | cc}
    \toprule
    \textbf{Model} & \textbf{NFEs} & \textbf{WER($\downarrow$)} & \textbf{UTMOS($\uparrow$)} & \textbf{SECS($\uparrow$)} & \textbf{FAD($\downarrow$)} & \textbf{CLAP($\uparrow$)} \\

    \midrule
     Ground Truth & -  & 2.21 & 4.00 & 0.9218 &  -   & 0.512   \\
    \midrule
     VoiceLDM & 200 & 14.01 & 2.82 & 0.7601 & 8.71 & 0.288 \\
     VoiceDiT & 200 & 11.08 & \textbf{3.33} & \textbf{0.8942}& 8.23 & 0.302 \\
     \midrule
     ImmersiveTTS & 25 & \textbf{9.89} & 3.23 & 0.8859& \textbf{7.81} & \textbf{0.323} \\
    \bottomrule
  \end{tabular}%
  } \vspace{-0.2cm}
    \caption{Objective evaluation results for single task on the LibriTTS \textit{test} and AudioCaps \textit{test} set.}\vspace{-0.3cm}
    \label{table:singletask}
\end{table}

\subsection{Single-Task Evaluation Results}
In addition to our main evaluation, we provide single-task results on TTS and TTA.
We additionally use UTMOS \citep{saeki2022utmos} as an objective proxy for perceived speech naturalness.
As shown in Table~\ref{table:singletask}, ImmersiveTTS yields the lowest WER among the diffusion baselines, suggesting improved intelligibility under the same evaluation setting.
In terms of speech naturalness, our system achieves UTMOS comparable to VoiceDiT while using substantially fewer sampling steps.
The SECS results show that ImmersiveTTS preserves speaker identity better than VoiceLDM, although it remains slightly below VoiceDiT.

For TTA, ImmersiveTTS attains the best FAD among the baselines and achieves the highest CLAP score, which is closest to \textit{Ground Truth}, indicating stronger text-audio semantic alignment.
We attribute this gain to our MM-DiT structure, which incorporates a dual-granularity conditioning strategy and an audio domain alignment objective.

%%%%%%%%%%%%%%%%%%%%%%%%%%%%%%%%%%%
%            TABLE 3
%%%%%%%%%%%%%%%%%%%%%%%%%%%%%%%%%%%
\begin{table}[t]
  \centering
  \resizebox{\columnwidth}{!}{%
  \begin{tabular}{l c cc c cc}
    \toprule
    & & \multicolumn{2}{c}{\textbf{Domain}} & \multicolumn{1}{c}{\textbf{Speech}} & \multicolumn{2}{c}{\textbf{Environment}} \\
    \cmidrule(lr){3-4} \cmidrule(lr){5-5} \cmidrule(lr){6-7}
    \textbf{Strategy} & \textbf{Target SSL} & \textbf{Speech} & \textbf{Env.} & \textbf{WER($\downarrow$)} & \textbf{FAD($\downarrow$)} & \textbf{CLAP($\uparrow$)} \\

    \midrule
    Base & - & - & -  & 11.21 & 9.64   & 0.236   \\
    
    \midrule
    \multirow{3}{*}{Single} 
     & WavLM & \ding{51} & - & 10.97 & 8.02 & 0.231 \\
     & ATST & - & \ding{51} & 13.77 & 8.78 & 0.271 \\
     & USAD & \ding{51} & \ding{51} & 9.04 & 7.93 & 0.239 \\
    \midrule
    \multirow{3}{*}{Dual}
     & WavLM, USAD & \ding{51} & \ding{51} & 8.95 & 7.33 & 0.248 \\
     & USAD, ATST & \ding{51} & \ding{51} & 8.94 & 8.20 & 0.266 \\
     & WavLM, ATST & \ding{51} & \ding{51} & \textbf{8.06} & \textbf{5.80} & \textbf{0.308} \\
    \bottomrule
  \end{tabular}%
  } \vspace{-0.2cm}
    \caption{Experimental results across different teacher alignment strategies.}\vspace{-0.25cm}
  \label{table:repamean}
\end{table}

\subsection{Analysis on Representation Alignment Strategy}
\label{ssec:repa_results}
To evaluate the effect of domain-specific REPA, we experiment with six teacher configurations: three single-teacher settings and three dual-teacher combinations.
We use the two encoders adopted in our model, WavLM and ATST-Frame, and additionally include USAD \citep{chang2025usad}, a unified speech-audio SSL encoder trained via distillation from these teachers.
\textit{Base} denotes our model trained without the REPA objective.
We follow the same experimental setup as in Section~\ref{ssec:main} and report results on the AudioCaps \textit{test} set.

We first examine the isolated effect of each teacher using a single-teacher strategy, as shown in Table~\ref{table:repamean}.
Using WavLM as the teacher yields a lower WER than the \textit{Base} and ATST-Frame-only setting, demonstrating improved content fidelity driven by the speech-focused target.
Although ATST-Frame degrades in intelligibility, it improves environment-related metrics, achieving the highest CLAP score and improving FAD over the \textit{Base}, suggesting better alignment with environmental context.
Because FAD is measured on the mixed waveform, stronger prompt alignment does not always translate into lower FAD, and we observe that WavLM tends to achieve lower FAD than ATST-Frame in this setting.
USAD improves all three metrics over the \textit{Base}, suggesting the benefit of guidance that covers both speech and environmental audio.

Based on these observations, we use a dual-teacher strategy and compare variants that include USAD with the domain-specific pair.
As shown in Table~\ref{table:repamean}, dual-teacher alignment mitigates the domain trade-off often observed in the single-teacher strategy.
Notably, the pairing of WavLM and ATST-Frame achieves the best performance across all metrics, outperforming configurations that incorporate USAD.
This indicates that the benefit of dual-teacher REPA comes not only from adding an additional target but also from choosing teachers with complementary strengths that provide more targeted guidance for speech content and environmental acoustics.
Additional experiments are provided in Appendix~\ref{appendix:repa}.
\begin{figure}[!t]
  \centering
\centerline{\includegraphics[width=0.5\textwidth]{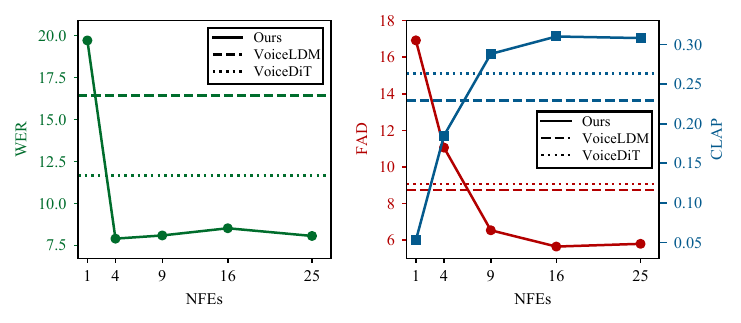}}\vspace{-0.4 cm}
\caption{Comparison across different NFEs.}
\label{fig:sampling} \vspace{-0.2cm}
\end{figure}
\subsection{Analysis on Sampling Steps}
We analyze how the number of sampling steps affects performance in environment-aware TTS.
We follow the same experimental setting as in Section~\ref{ssec:main} with AudioCaps \textit{test} set.
Figure~\ref{fig:sampling} shows consistent improvements as the sampling step increases.
Increasing NFEs reduces both WER and FAD, while improving CLAP score, with the largest gains observed when moving from very few steps to moderate steps (e.g., $1\to9$ steps).

Notably, ImmersiveTTS matches or exceeds the diffusion baselines with far fewer sampling steps. 
With only 9 steps, it achieves lower WER and FAD, and higher CLAP scores than VoiceLDM and VoiceDiT, both of which use 200 NFEs. 
This highlights a favorable quality-efficiency trade-off, delivering comparable or better quality with substantially fewer inference steps.

\section{Conclusion}
We presented ImmersiveTTS, an environment-aware text-to-speech framework that jointly synthesizes intelligible speech and environmental audio within a unified flow matching-based generative model.
Built on an MM-DiT backbone, ImmersiveTTS explicitly models cross-modal interactions through a dual-stream stage that fuses transcript-aligned speech features with text-conditioned environmental cues via joint attention.
To mitigate the domain mismatch between speech and environmental audio and to stabilize training, we further introduce a domain-specific REPA objective that aligns intermediate representations with distinct SSL teachers specialized for speech and environmental audio, respectively.
Across evaluations on real and augmented environment-aware TTS benchmarks, ImmersiveTTS yields higher overall quality and semantic fidelity than existing approaches.
Comprehensive analysis and ablation studies confirmed the effectiveness of the proposed dual-stream interaction design and domain-specific REPA for environment-aware TTS.

\section{Limitations}
Despite the effectiveness of ImmersiveTTS in jointly synthesizing speech with environmental audio, we acknowledge several limitations. 
First, our training relies primarily on synthetic mixtures of speech and environmental audio. 
While this enables scalable training, it may not fully capture the complex acoustic interactions present in large-scale recordings in the wild.
We also note that robustness across varying SNR conditions and scene difficulty levels remains underexplored in the current work.
While ImmersiveTTS ensures robust control over linguistic content and speaker identity through its dedicated modules, it currently lacks explicit control over paralinguistic attributes such as prosody, speaking style, or emotion. 
Incorporating these factors will be a crucial next step for producing expressive speech that fully aligns with both the target scene and the intended emotional expression.
In future work, we aim to address these limitations by exploring large-scale real-world recordings and developing a method for the granular control of paralinguistic factors to achieve more immersive speech synthesis.

\section{Acknowledgments}
This work was partly supported by Institute of Information \& Communications Technology Planning \& Evaluation (IITP) under the artificial intelligence graduate school program (Korea University) (No. RS-2019-II190079) and artificial intelligence star fellowship support program to nurture the best talents (IITP-2026-RS-2025-02304828) grant funded by the Korea government (MSIT).

\bibliography{custom}
\cleardoublepage
\appendix

\section{Additional Implementation Details}
\label{appendix:implementation}
\begin{figure}[!t]
  \centering
\centerline{\includegraphics[width=0.5\textwidth]{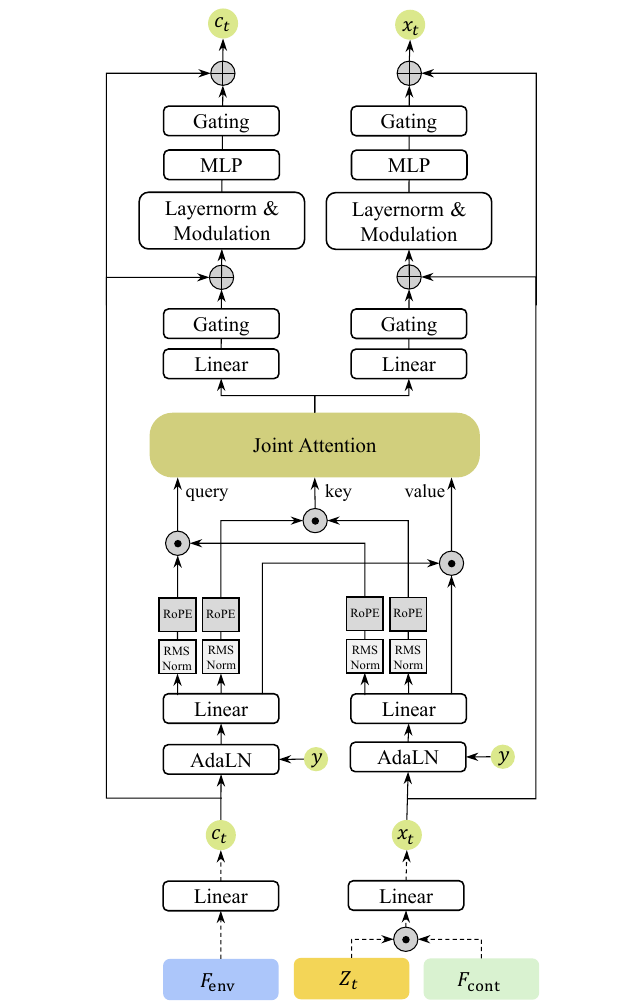}}
\caption{Illustration of the double-stream DiT blocks. Dashed lines indicate that this input initialization is applied only to the first double-stream block, where $F_{\mathrm{env}}$ denotes the Flan-T5 feature for the environment prompt, $Z_t$ denotes the noisy latent at timestep $t$, and $F_{\mathrm{cont}}$ denotes the content-conditioned feature derived from the content encoder.}
\label{fig:mmditblock} \vspace{-0.2cm}
\end{figure}

To preserve speaker identity, we extract a speaker embedding from the speaker prompt using a pretrained WavLM-based speaker verification model\footnote{\url{https://huggingface.co/microsoft/wavlm-base-sv}}.
This embedding is projected and fed as an additional conditioning input to the text encoder during both training and inference.   
For the environment prompt, we utilize the last hidden states of Flan-T5-\textit{Large}\footnote{\url{https://huggingface.co/google/flan-t5-large}} \citep{chung2024scaling} as the token sequence for the environment context stream, while the global output vector of CLAP\footnote{\url{https://huggingface.co/laion/clap-htsat-unfused}} provides coarse conditioning features.
Figure~\ref{fig:mmditblock} provides a zoomed-in view of the internal flow of the double-stream DiT block.

% [Inference Setup]
During inference, we set the classifier-free guidance scales to $\omega_{\text{env}} = 3$ and $\omega_{\text{cont}} = 3$ for each sub-modality.
For the vocoder, we use the pretrained HiFi-GAN \citep{kong2020hifi} to reconstruct the 16 kHz waveform from the sampled mel-spectrogram. 

For representation alignment, we extract target features using WavLM-\textit{Large}\footnote{\url{https://huggingface.co/microsoft/wavlm-large}}, ATST-Frame-\textit{Base}\footnote{\url{https://github.com/Audio-WestlakeU/audiossl}}, and USAD-\textit{Base}\footnote{\url{https://huggingface.co/MIT-SLS/USAD-Base}}.
We use the representations from the final layer of each encoder.
Crucially, WavLM operates on clean speech from LibriTTS before mixing, ensuring that its alignment target focuses solely on linguistic fidelity.
In contrast, ATST-Frame operates on mixed speech with environmental audio, allowing the target to capture the full acoustic scene. 
All teacher encoders remain frozen during training.

\begin{figure*}[t]
  \centering
  \includegraphics[width=0.75\textwidth]{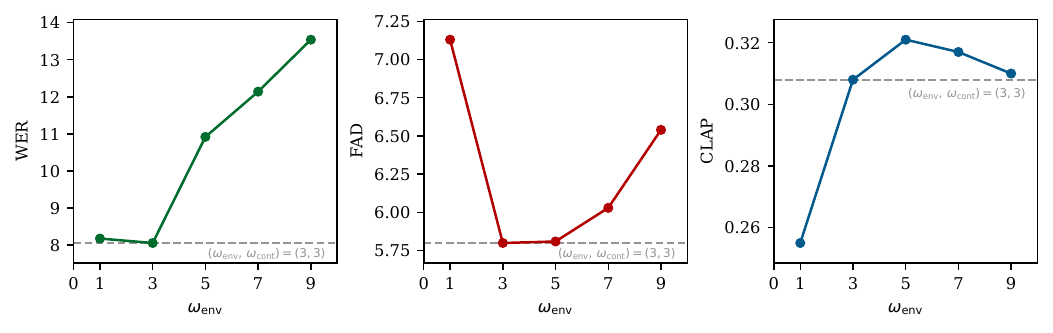} 
  \vspace{-0.2cm}
  \caption {Objective evaluation results on AudioCaps \textit{test} set across various environment guidance scales.} \vspace{-0.2cm}
  \label{fig:cfg_env}
\end{figure*}

\begin{figure*}[t]
  \centering
  \includegraphics[width=0.75\textwidth]{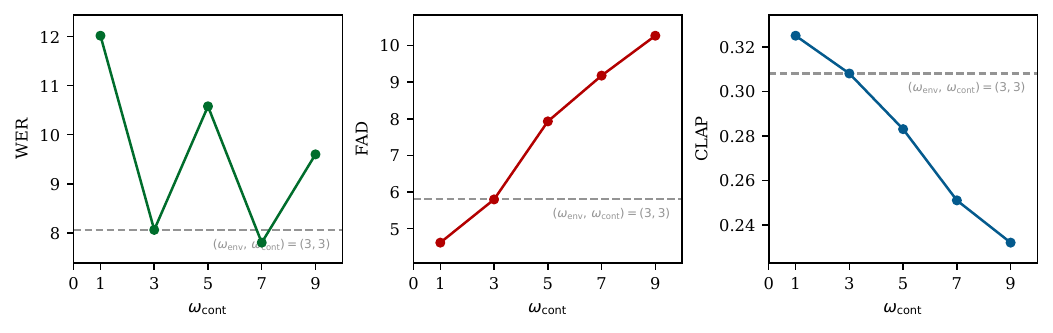} 
  \vspace{-0.2cm}
  \caption {Objective evaluation results on AudioCaps \textit{test} set across various content guidance scales.} \vspace{-0.2cm}
  \label{fig:cfg_cont}
\end{figure*}

\section{Details of Subjective Evaluations}
\label{appendix:mos}
For the subjective evaluation, we conducted a mean opinion score (MOS) test to assess four aspects of the generated audio on a 5-point scale (1 to 5):
speech naturalness (SN-MOS), environmental consistency (EC-MOS), overall integration naturalness (ON-MOS), and speaker similarity (S-MOS).
SN-MOS measures the perceived naturalness of the synthesized speech; EC-MOS assesses how well the background audio matches the given environment description; ON-MOS evaluates overall naturalness, i.e., how naturally the speech and background audio are blended; and S-MOS measures how similar the synthesized speech is to the reference speaker in terms of speaker identity.
To construct the evaluation data, we randomly sampled 30 utterances from each test set, excluding samples whose ground-truth audio contains multiple speakers. 
We didn't include the ground truth samples for the subjective evaluation, as they are perceptually very similar to the reconstructed samples produced by VAE and vocoder, which could lead to redundant comparisons and potentially bias listeners. 
All MOS ratings are reported with 95\% confidence intervals.

\begin{figure*}[!t]
  \centering
\centerline{\includegraphics[width=1.0\textwidth]{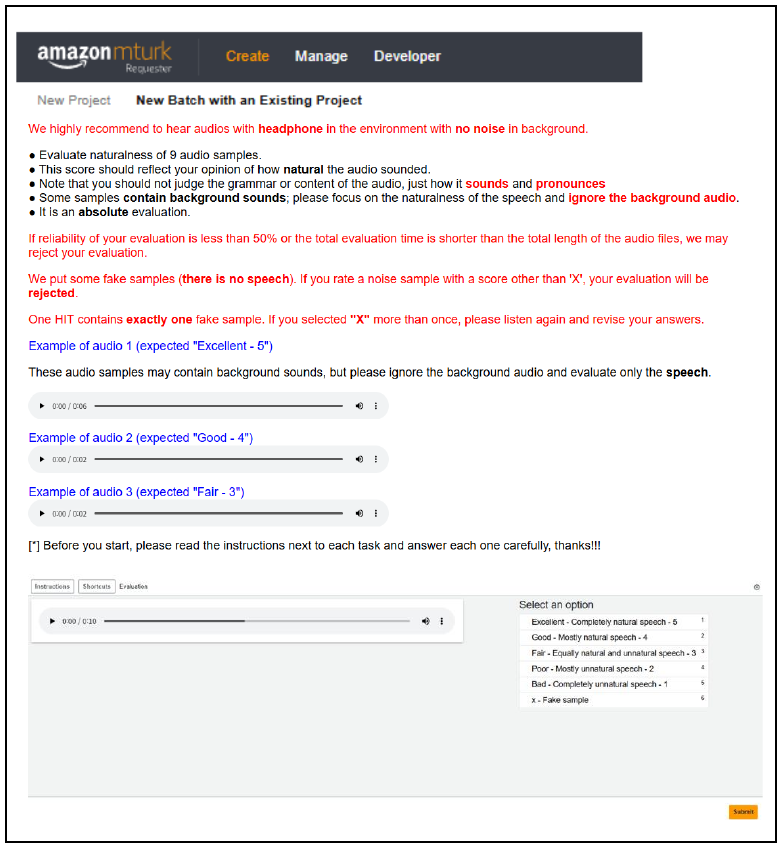}}\vspace{-0.25 cm}
\caption{Detailed information on listener requirements and the SN-MOS evaluation interfaces.}
\label{fig:sn_mos} \vspace{0.0cm}
\end{figure*}

\begin{figure*}[!t]
  \centering
\centerline{\includegraphics[width=1.0\textwidth]{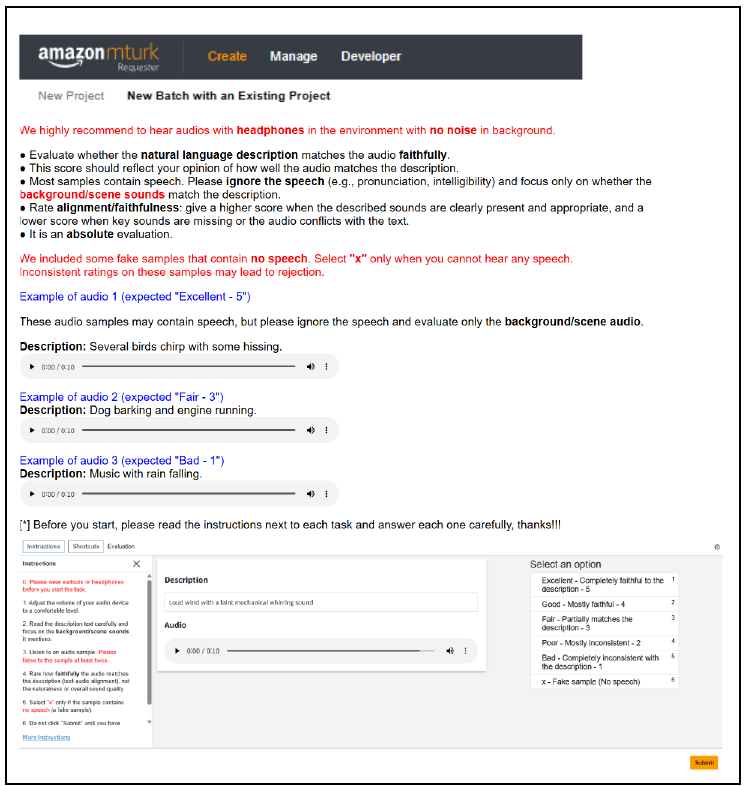}}\vspace{-0.25 cm}
\caption{Detailed information on listener requirements and EC-MOS evaluation interfaces.}
\label{fig:ec_mos} \vspace{0.0cm}
\end{figure*}

\begin{figure*}[!t]
  \centering
\centerline{\includegraphics[width=1.0\textwidth]{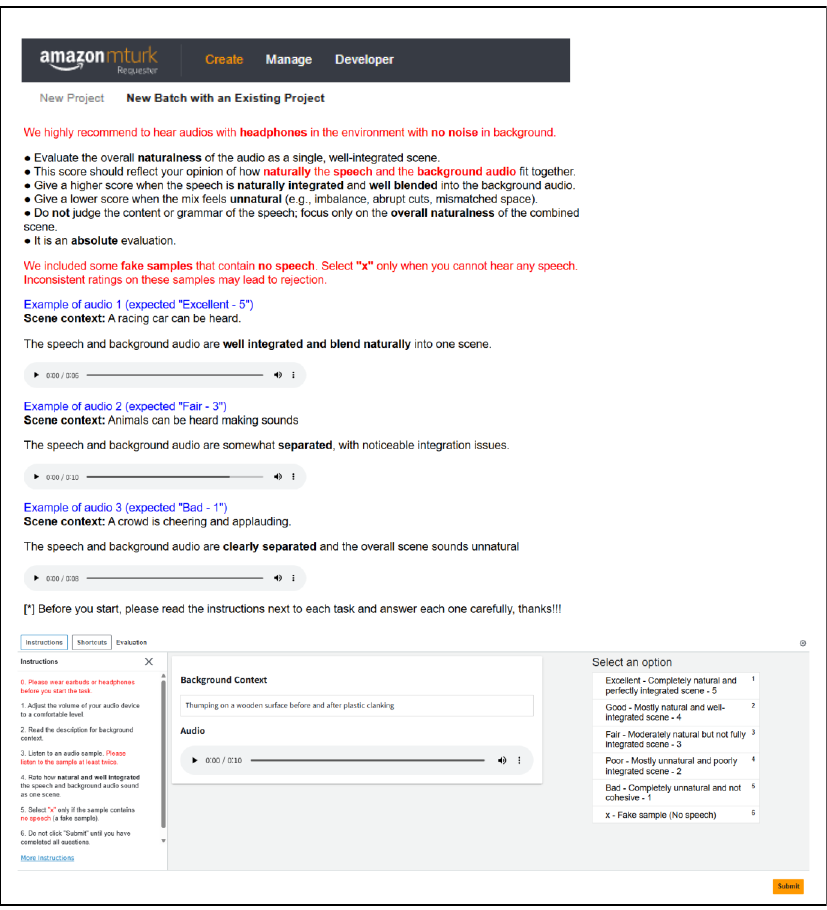}}\vspace{-0.25 cm}
\caption{Detailed information on listener requirements and ON-MOS evaluation interfaces.}
\label{fig:on_mos} \vspace{0.0cm}
\end{figure*}

\begin{figure*}[!t]
  \centering
\centerline{\includegraphics[width=1.0\textwidth]{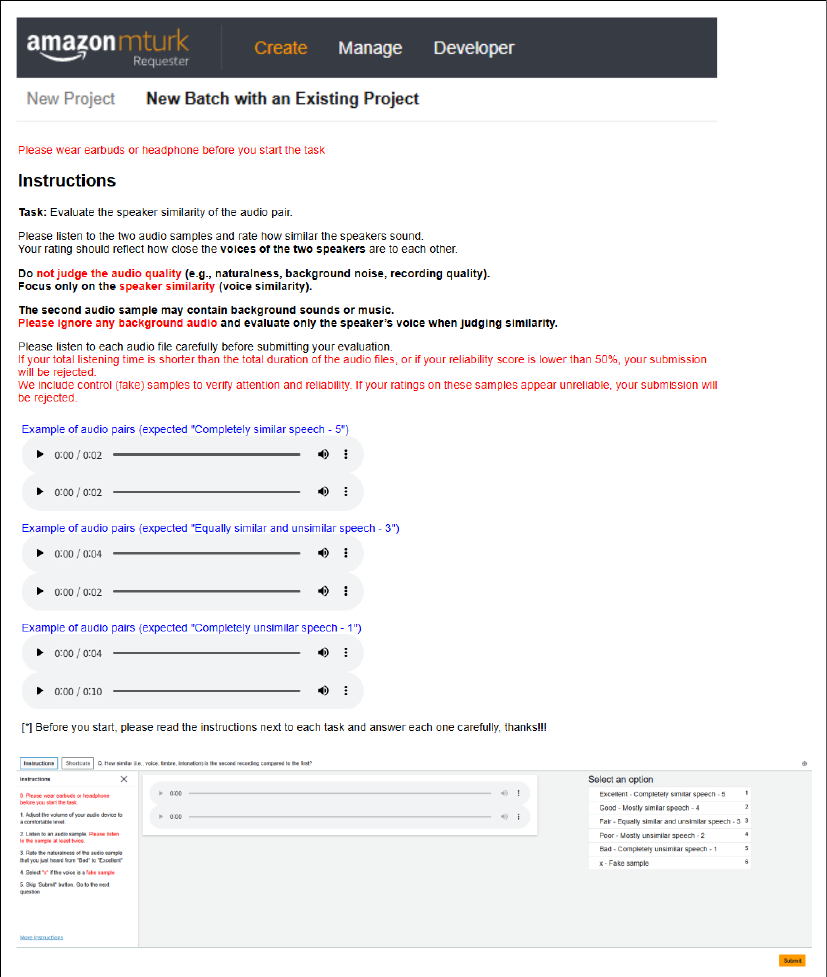}}\vspace{-0.25 cm}
\caption{Detailed information on listener requirements and S-MOS evaluation interfaces.}
\label{fig:s_mos} \vspace{0.0cm}
\end{figure*}

We conducted these evaluations via crowdsourcing on Amazon Mechanical Turk\footnote{\url{https://www.mturk.com/}}.
Each evaluation was completed by 20 native English speakers residing in the United States.
We allocated 42 USD per MOS task and ran the SN-MOS, EC-MOS, and ON-MOS assessments separately for each test set. 
In addition, the S-MOS evaluation on one test set costs 84 USD, resulting in a total cost of 336 USD.
We additionally interspersed fake samples as attention checks throughout each evaluation set.
We excluded ratings from participants who failed these checks.
Screenshots of the Amazon Mechanical Turk interface for SN-MOS, EC-MOS, ON-MOS, S-MOS are shown in Figure~\ref{fig:sn_mos}, ~\ref{fig:ec_mos}, ~\ref{fig:on_mos}, and ~\ref{fig:s_mos}.

\begin{table}[t]
  \centering
  \resizebox{\columnwidth}{!}{%
  \begin{tabular}{c c c cc}
    \toprule
     \textbf{Alignment Position}    & \textbf{Target SSL} & \textbf{WER($\downarrow$)} & \textbf{FAD($\downarrow$)} & \textbf{CLAP($\uparrow$)} \\ \midrule
    \multicolumn{5}{c}{\textbf{Single Teacher}} \\ \midrule
    
    \multirow{3}{*}{MM-DiT}
    & WavLM  & 10.97 & 8.02   & 0.231 \\ 
    & ATST   & 13.77 & 8.78   & 0.271 \\ 
    & USAD   & 9.04 & 7.93   & 0.239 \\ 
    \midrule
    \multirow{3}{*}{Single DiT}
    & WavLM   & 9.94 & 7.83   & 0.226 \\ 
    & ATST   & 14.57 & 7.24   & 0.261 \\ 
    & USAD   & 10.47 & 7.02   & 0.266 \\ 
    \midrule

    \multicolumn{5}{c}{\textbf{Dual-Teacher}} \\

    \midrule
    \multirow{3}{*}{MM-DiT \& Single DiT}
    & WavLM + USAD   & 9.31 & 7.76   & 0.231 \\ 
    & USAD + ATST   & 12.42 & 8.66   & 0.247 \\ 
    & WavLM + ATST  & 9.67 & 5.97   & 0.287 \\ 
    \midrule
    
    \multirow{3}{*}{MM-DiT \& MM-DiT}
    &   WavLM + USAD   & 8.95 & 7.33   & 0.248 \\ 
    &  USAD + ATST   & 8.94 &  8.20 & 0.266 \\ 
    &  WavLM + ATST   & \textbf{8.06} & \textbf{5.80}   & \textbf{0.308} \\ 
    \bottomrule
  \end{tabular}%
  }
    \vspace{-0.2cm}\caption{Experimental results across different teachers and alignment positions on the AudioCaps \textit{test} set.}\vspace{-0.2cm}
  \label{table:repa_position} 
\end{table}

\section{Analysis on Representation Alignment Injection Position}
\label{appendix:repa}
In this section, we report the effect of REPA injection position. 
As shown in Section~\ref{ssec:repa_results}, ImmersiveTTS establishes the benefit of complementary teachers.
We check whether the same trend holds when the alignment loss is applied at different positions within the DiT backbone.

In our preliminary experiments, aligning WavLM at ten locations across the 30 DiT blocks did not reveal a clear monotonic pattern.
However, aligning in the middle (or slightly earlier) layers was consistently more stable, consistent with prior findings \citep{yu2024representation}.
Based on this observation, we focus on the specific injection points: the $6$th or $10$th block in the MM-DiT stage (12 blocks) and the $4$th block in the single DiT stage (18 blocks).

Table~\ref{table:repa_position} reports objective results on the AudioCaps \textit{test} set under different injection positions and teacher configurations.
Overall, the qualitative trends observed in Section~\ref{ssec:repa_results} remain consistent across the injection stage.
In the single teacher setting, WavLM primarily improves speech accuracy, whereas ATST-Frame improves environment-related metrics, indicating that each teacher provides domain-specific semantic guidance. 
USAD exhibits a more balanced behavior depending on the injection stage. 
In the dual-teacher setting, combining WavLM and ATST-Frame consistently yields the best overall performance, and injecting both alignments within the MM-DiT blocks achieves the strongest results, suggesting robustness to this design choice.

\begin{table}[t]
  \centering
  \resizebox{\columnwidth}{!}{%
  \begin{tabular}{c c c c c c}
    \toprule
     \textbf{Model} & Reconstructed & VoiceLDM  & VoiceDiT & ImmersiveTTS  \\ \midrule
    \textbf{S-MOS($\uparrow$)}  & 3.18 $\pm$  0.05 &  2.97 $\pm$  0.06 & \textbf{3.15} $\pm$  0.06 & \textbf{3.15} $\pm$  0.06 \\
    \bottomrule
  \end{tabular}%
  } \vspace{-0.2cm}
    \caption{S-MOS results for environment-aware TTS on the Seed-TTS \textit{test-en} and AudioCaps \textit{test} sets.}\vspace{-0.25cm}
        \label{table:spk_smos}
\end{table}

\begin{table}[t]
  \centering
  \resizebox{\columnwidth}{!}{%
  \begin{tabular}{l | c | cc | cc}
    \toprule
    \textbf{Model} & \textbf{NFEs} & \textbf{WER($\downarrow$)} & \textbf{SECS($\uparrow$)} & \textbf{FAD($\downarrow$)} & \textbf{CLAP($\uparrow$)} \\

    \midrule
     Ground Truth & -  & 2.21 & 0.9218 &  -   & 0.512   \\
    \midrule
     VoiceLDM & 200 & 14.01 & 0.7601 & 8.71 & 0.288 \\
     VoiceDiT & 200 & 11.08 & 0.8942 & 8.23 & 0.302 \\
     \midrule
     CosyVoice2 & 10 & \textbf{5.23} & \textbf{0.9227} & - & - \\ 
     CosyVoice3 & 10 & 5.98 & 0.9152 & - & - \\
     \midrule
     AudioLDM2 (Audio) & 200 & - & - & 3.79 & 0.407 \\ 
     TangoFlux  & 25 & - & - & \textbf{2.96} & \textbf{0.502}  \\ 
     \midrule
     ImmersiveTTS & 25 & 9.89  & 0.8859& 7.81 & 0.323 \\
    \bottomrule
  \end{tabular}%
  } \vspace{-0.2cm}
    \caption{Objective evaluation results for single task on the LibriTTS \textit{test} and AudioCaps \textit{test} set with single task baselines.}\vspace{-0.25cm}
      \label{table:appendix_singletask}
\end{table}

\section{Analysis on Dual Classifier-Free Guidance Scale}
To independently control sub-modality attributes, ImmersiveTTS adopts dual classifier-free guidance (CFG) with separate guidance scales for the environmental condition and the content condition.
In the main experiments (Section~\ref{ssec:main}), we use $(\omega_{\text{env}}, \omega_{\text{cont}})=(3,3)$.
Here, to analyze the sensitivity to each guidance scale, we fix one scale to $3$ and vary the other in $\{1,3,5,7,9\}$, reporting WER, FAD, and CLAP on the same evaluation setting.

Figure~\ref{fig:cfg_env} varies $\omega_{\text{env}}$ with $\omega_{\text{cont}}=3$.
We observe that increasing $\omega_{\text{env}}$ beyond $3$ substantially degrades intelligibility.
WER increases from $8.06$ at $\omega_{\text{env}}=3$ to over $10.92$ for $\omega_{\text{env}}\ge 5$. 
While FAD exhibits only mild variations, and CLAP shows a small improvement at moderate $\omega_{\text{env}}$ before plateauing at higher values.
This suggests that overly strong environmental guidance can interfere with linguistic realization, even if it slightly improves text-audio alignment for the background.
Conversely, setting $\omega_{\text{env}}=1$ reduces semantic alignment and yields worse perceptual quality than $\omega_{\text{env}}=3$.

Figure~\ref{fig:cfg_cont} varies $\omega_{\text{cont}}$ while fixing $\omega_{\text{env}}=3$.
The lowest WER is achieved at $\omega_{\text{cont}}=7$.
In contrast, as $\omega_{\text{cont}}$ increases, acoustic realism degrades steadily, reflected by a monotonic increase in FAD from $4.62$ to $10.27$.
At the same time, semantic alignment decreases consistently, as CLAP drops from 0.325 to 0.232 over the same range.
Overall, these results suggest that overly large $\omega_{\text{cont}}$ over-emphasizes speech content, improving intelligibility only up to a point while harming overall audio quality and scene coherence.

Overall, the analysis reveals a trade-off between speech clarity and semantic quality when scaling dual CFG.
The balanced setting $(\omega_{\text{env}}, \omega_{\text{cont}})=(3,3)$ used in our main experiments provides a stable operating point, avoiding the sharp WER degradation observed with large $\omega_{\text{env}}$, and the FAD and CLAP collapse observed with large $\omega_{\text{cont}}$.

\section{Speaker Identity Preservation Evaluation Results}
\label{appendix:smos}

To further assess speaker identity preservation in the main environment-aware TTS setting, we conduct a S-MOS evaluation on the same augmented test set used in Table~\ref{table:envtts2}, where clean speech from Seed-TTS \textit{test-en} is mixed with non-speech environmental audio from AudioCaps \textit{test}. 
For each sample, we use the corresponding clean target speech waveform from Seed-TTS \textit{test-en} as the reference speech for speaker similarity evaluation.
As shown in Table~\ref{table:spk_smos}, ImmersiveTTS achieves an S-MOS of 3.15, outperforming VoiceLDM and matching VoiceDiT. 
This score is also close to that of the reconstructed samples (3.18), suggesting that speaker identity is largely preserved in the environment-aware TTS setting.

\begin{table}[t]
  \centering
  \resizebox{\columnwidth}{!}{%
  \begin{tabular}{l c cc}
    \toprule
     \textbf{Model}     & \textbf{WER($\downarrow$)} & \textbf{FAD($\downarrow$)} & \textbf{CLAP($\uparrow$)} \\ \midrule
    
    VoiceLDM  & 16.45 & 8.75   & 0.229 \\ 
    VoiceDiT  & 11.68 & 9.07   & 0.263 \\ 
    AudioLDM2 (Speech) & 35.06 & 29.59 & 0.048 \\ 
    \midrule
    AudioLDM2 (Speech) + AudioLDM2 (Audio) & 41.33 & 5.36 & 0.365 \\ 
    CosyVoice2 + TangoFlux  & \textbf{6.76} & \textbf{4.01} & \textbf{0.452} \\ 
    \midrule
    ImmersiveTTS  & 8.06 & 5.80 & 0.308 \\ 
    \bottomrule
  \end{tabular}%
  } \vspace{-0.2cm}
    \caption{Objective results for environment-aware TTS on the AudioCaps \textit{test} set. The symbol ‘+’ denotes mixing of separately generated speech and background audio.}\vspace{-0.2cm}
  \label{table:appendix_mixing}
\end{table}

\begin{table}[t]
  \centering
  \resizebox{\columnwidth}{!}{%
  \begin{tabular}{l c cc}
    \toprule
     \textbf{Model}     & \textbf{WER($\downarrow$)} & \textbf{FAD($\downarrow$)} & \textbf{CLAP($\uparrow$)} \\ \midrule
    
    VoiceLDM  & 11.20 & 6.98 & 0.118 \\ 
    VoiceDiT  & 7.08 & 5.37 & 0.134 \\ 
    AudioLDM2 (Speech) & 27.23 & 28.14 & -0.069 \\ 
    \midrule
    AudioLDM2 (Speech) + AudioLDM2 (Audio) & 35.78 & 3.39 & 0.236 \\ 
    CosyVoice2 + TangoFlux  & \textbf{2.28} & \textbf{3.11} & \textbf{0.287} \\ 
    \midrule
    ImmersiveTTS  & 4.48 & 3.92 & 0.207 \\ 
    \bottomrule
  \end{tabular}%
  } \vspace{-0.2cm}
    \caption{Objective results for environment-aware TTS on the Seed-TTS \textit{test-en} and AudioCaps \textit{test} sets. The symbol ‘+’ denotes mixing of separately generated speech and background audio.}\vspace{-0.2cm}
    \label{table:appendix_mixing2}
\end{table}

\section{Broader Baseline Comparisons}
\label{appendix:broad_baseline}
\subsection{Single-Task Baselines}
To provide a broader comparison beyond the environment-aware TTS baselines in the main paper, we compare against CosyVoice2 \citep{du2024cosyvoice} and CosyVoice3 \citep{du2025cosyvoice} for TTS, and against AudioLDM2-\textit{Audio} \citep{liu2024audioldm} and TangoFlux \citep{hung2024tangoflux} for TTA under the same evaluation settings as Table~\ref{table:singletask}. 
Table~\ref{table:appendix_singletask} summarizes the results. 
As expected, single-task models perform strongly on their respective metrics. 
For TTS, CosyVoice2 and CosyVoice3 achieve substantially lower WER and higher SECS than the environment-aware TTS models, while for TTA, AudioLDM2 and TangoFlux achieve markedly better FAD and CLAP. 
These results reflect the inherent trade-off in environment-aware TTS, as a unified model must balance speech fidelity and background generation within a single system.

\subsection{Mixing-based Pipeline Baselines}
Building on the single-task baselines above, we further construct mixing-based pipeline baselines by separately generating speech and background audio and then mixing the two outputs.
We consider two mixing-based baselines: one separately generates speech and audio using domain-specific AudioLDM2 checkpoints, and the other combines CosyVoice2 for speech with TangoFlux for background audio.
Table~\ref{table:appendix_mixing} follows the same setting as Table~\ref{table:envtts}, while Table~\ref{table:appendix_mixing2} follows the augmented setting used in Table~\ref{table:envtts2}.
As shown in both Table~\ref{table:appendix_mixing} and Table~\ref{table:appendix_mixing2}, AudioLDM2-based pipelines show relatively poor WER, mainly because the speech generated by AudioLDM2-\textit{Speech} is less intelligible.
The combination with CosyVoice2 and TangoFlux consistently performs the strongest across all objective metrics in both settings.

In contrast, such pipelines require separate generation and mixing, whereas ImmersiveTTS directly models speech-background interaction within a unified framework.
We believe this unified formulation is important for improving coherence and realism by modeling how speech and background influence each other during synthesis.

\section{Potential Risks}
The proposed environment-aware TTS system is designed to synthesize speech together with environmental audio based on provided textual prompt. As with other speech generation models, this capability inherently entails several potential risks. The system could be misused to generate unauthorized voice synthesis or deceptive audio content, potentially causing negative societal impact.

To mitigate these potential risks, our work is intended solely for research purposes, and we emphasized the importance of transparent disclosure of synthesized content and responsible use. Furthermore, when releasing our resources, we explicitly encourage users to adhere to these principles. 
\section{AI Assistance}
We used ChatGPT 5.2 to assist with proofreading and improving English grammar and expressions.

\end{document}